\documentclass[conference, 10pt]{IEEEtran}
\ifCLASSINFOpdf
\else
\fi
\usepackage{graphicx}
\usepackage{color}
\usepackage{placeins}
\usepackage{float}
\usepackage{hyperref}
\usepackage{tabularx,colortbl}

\usepackage{tikz}
\usepackage{textcomp}
\usepackage{hyperref}
\usepackage{lipsum}

\newcommand\copyrighttext{%
  \footnotesize \textcopyright 2017 IEEE. Personal use of this material is permitted.
  Permission from IEEE must be obtained for all other uses, in any current or future
  media, including reprinting/republishing this material for advertising or promotional
  purposes, creating new collective works, for resale or redistribution to servers or
  lists, or reuse of any copyrighted component of this work in other works.
  }
\newcommand\copyrightnotice{%
\begin{tikzpicture}[remember picture,overlay]
\node[anchor=south,yshift=10pt] at (current page.south) {\fbox{\parbox{\dimexpr\textwidth-\fboxsep-\fboxrule\relax}{\copyrighttext}}};
\end{tikzpicture}%
}

\begin{document}
%

\title{Silicon Micromachined High-contrast Artificial Dielectrics for Millimeter-wave Transformation Optics Antennas}


\author{\IEEEauthorblockN{Nicolas Garcia, Wenlong Bai, Thibault Twahirwa, David Connelly and Jonathan Chisum}
\IEEEauthorblockA{Department of Electrical Engineering\\
University of Notre Dame\\
Notre Dame, IN, USA\\
ngarcia7@nd.edu, Wenlong.Bai.8@nd.edu, ttwahirw@nd.edu, dconnel7@nd.edu, jchisum@nd.edu\\}}


%


\maketitle
\copyrightnotice

\begin{abstract}
Transformation optics methods and gradient index electromagnetic structures rely upon spatially varied arbitrary permittivity. This, along with recent interest in millimeter-wave lens-based antennas demands high spatial resolution dielectric variation. Perforated media have been used to fabricate gradient index structures from microwaves to THz but are often limited in contrast. We show that by employing regular polygon unit-cells (hexagon, square, and triangle) on matched lattices we can realize very high contrast permittivity ranging from 0.1--1.0 of the background permittivity. Silicon micromachining (Bosch process) is performed on high resistivity Silicon wafers to achieve a minimum permittivity of 1.25 (10\% of Silicon) in the WR28 waveguide band, specifically targeting the proposed $39\,$GHz 5G communications band. The method is valid into the THz band.
\end{abstract}



%
\IEEEpeerreviewmaketitle

\section{Introduction}

Recent advances in transformation optics (TO) have reinvigorated investigation into gradient index (GRIN) optics and in particular gradient index lenses and antennas \cite{kwon_transformationEM_2010}. Fabrication of GRIN lenses has been intensely studied from additive \cite{teichman_gradient_2013} and subtractive (traditional) manufacturing. A popular approach for spatially varying permittivity throughout a bulk material was proposed in \cite{potosa_perforateddielectric_1994} where they mechanically drilled circular voids in a background dielectric on regular square and triangular lattices. The effective permittivity of a given perforation, or void, on a lattice unit-cell is
\begin{equation}
    \epsilon_{\textrm{\tiny eff}} = \epsilon_r\left(1-\alpha\right)+\alpha,
\end{equation}
\noindent where $\epsilon_r$ is the relative permittivity of the background dielectric and $\alpha$ is the filling factor equal to the ratio of the void area to the unit-cell area. For circular unit-cells on square and triangle lattices the minimum fill factor is equal to $\frac{\pi}{4}$ and $\frac{\pi}{2\sqrt{3}}$, respectively. If the background permittivity is that of Silicon, $\epsilon_r=11.8$, the corresponding minimum effective permittivity is 3.3 and 2.0, respectively which limits the fabrication of many TO designs. For example, the permittivity of a flat lens varies from a maximum value (dependent on the thickness of the lens) to that of air \cite{kwon_transformationEM_2010}.

In this work we investigate Silicon micromachined perforated media for application in millimeter-wave TO and GRIN antenna designs. We use the Bosch Deep Reactive Ion Etch process in undoped, high-resistivity Silicon wafers to maximize the range of realizable permittivity and emphasize fill factors approaching $\alpha=1.0$ while maintaining manufacturability and remaining self supporting (note that $\alpha=1.0$ is a null wafer in which the Silicon has been completely etched away). Photolithographic drilling (versus mechanical drilling) permits arbitrary perforation cross sections on arbitrary lattices that also meet sub-wavelength unit-cell requirements at millimeter-waves. To approach the minimum permittivity of air we fabricated the set of all regular lattice $n$-gons ($n=3,4,\textrm{ and }6$ corresponding to triangles, squares, and hexagons) on their corresponding $n$-gon lattices. Each has the property that as the characteristic dimension of the void approaches that of the unit-cell, $\alpha=1$ and $\epsilon_{\textrm{\tiny eff}}=\epsilon_r$. 

\begin{figure}[t]
\begin{center}
\noindent
  \includegraphics[width=\columnwidth]{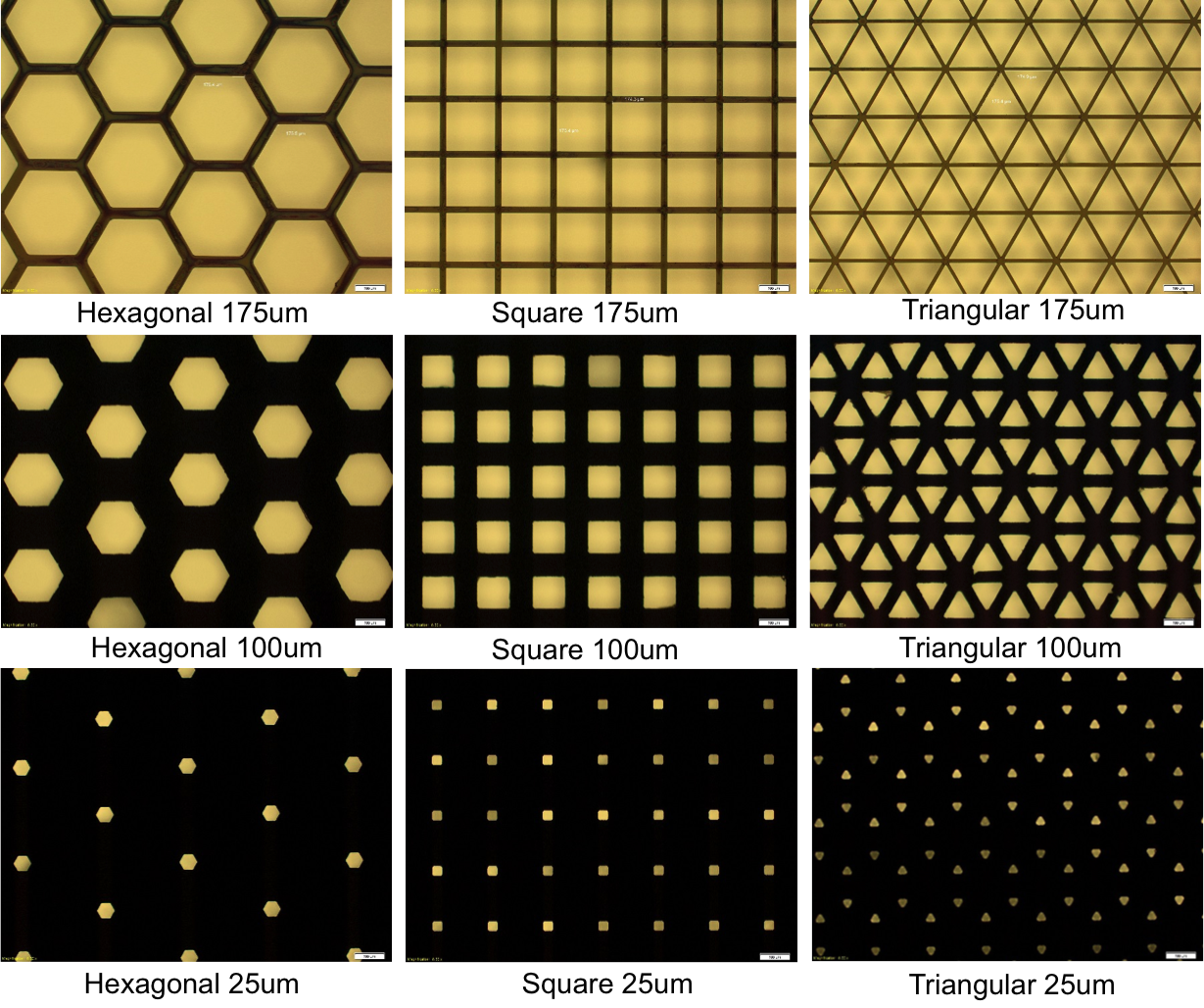}
  \caption{Hexagonal, square, and triangular perforations with characteristic dimension $175\mu$m, $100\mu$m, and $25\mu$m as seen with backlight at 10x magnification. All perforations are on a regular $200\mu$m lattice which is $\lambda_g/11$ at $39\,$GHz in bulk Silicon.}
  \label{fig:Fabrication}
\vspace{-10mm}
\end{center}
\end{figure}

To explore trade-offs in mechanical strength and manufacturability we fabricated the hexagon, square, and triangle features on a $200\mu$m lattice spacing ($\lambda_g/11$ at 39\,GHz) with characteristic dimensions $d=175\mu$m, $100\mu$m, and $25\mu$m as shown in Fig.\,\ref{fig:Fabrication} at $10\times$ magnification. The maximum feature size, $d=175\mu$m, has a supporting bridge of width $50\mu$m between each void. Comsol finite-element simulations of a $150\mu$m square perforation showed maximum surface stress to be much less than the yield strength of Silicon indicating that our $175\mu$m perforations should also be self supporting.

\section{Approach}

Each feature in Fig.\,\ref{fig:Fabrication} was exposed onto $25.4\,$mm, $280\mu$m thick undoped $\left<111\right>$ intrinsic float-zone Silicon wafers with $>2000 \Omega$-cm resistivity. The wafers were coated with $12\mu$m of 4620 photoresist (following the thick photoresist recipe). After development, the samples were placed in an Alcatel 601E inductively coupled reactive ion etcher and anisotropically etched with a DRIE Bosch process for 45, 60, and 120 minutes (for $175\mu$m, $100\mu$m, and $25\mu$m features, respectively). The $25\mu$m minimum feature size was chosen because of the mask machine used but in principle minimum voids could be arbitrarily small (sub-micron) and therefore maximum permittivity could be approximately that of the background permittivity. 

Each geometry was arrayed in $6.9\times3.4\,$mm rectangles corresponding to the inner dimensions of a WR28 rectangular waveguide (dimensions $7.112\times3.556\,$mm). Several rectangular samples were stacked in a quarter-wave section of WR28 rectangular waveguide and scattering parameters were measured across the waveguide band from $26.5-40\,$GHz. A thru-reflect-line (TRL) calibration was performed to set the reference plane at the entrance ports of the quarter-wave section and the Nicolson-Ross Weir method \cite{weir_material_1974} was used to extract the effective permittivity from scattering parameters. It was found that significant mismatches at the air-silicon interface caused poor measurement accuracy and so we augmented the Nicolson-Ross Weir method to include impedance matching sections in the waveguide. The complete material stack-up placed inside the waveguide consisted of three $0.2794\,$mm thick Silicon samples placed between two impedance matching Rogers 4350B substrate samples, each $0.7366\,$mm thick and with a measured dielectric constant of 3.7. Measurements were also highly sensitive to the air-gap between the top of the Silicon samples and the top of the waveguide so the 76$\mu$m gap was accounted for in the permittivity extraction but it should be noted that some error in the final effective permittivity is attributed to this extraction. The measurement setup was validated with samples of un-etched bulk Silicon with measured $\epsilon_r=12.0$.

\section{Results and Conclusion}

\begin{figure}[tb]
\begin{center}
\noindent
  \includegraphics[width=\columnwidth]{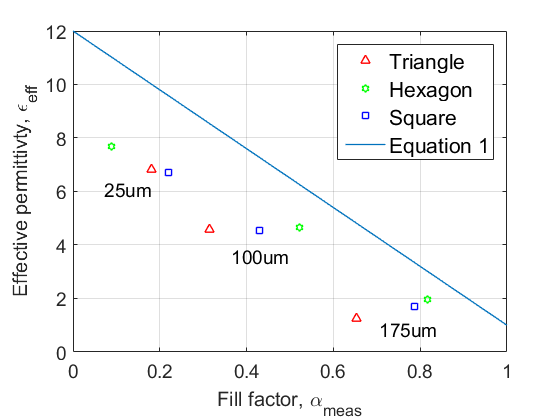}
  \caption{Average permittivity of each geometry at each feature size.}
  \label{fig:Results}
  \vspace{-8mm}
\end{center}
\end{figure}

Figure\,\ref{fig:Results} shows measured effective permittivity for each geometry as a function of measured fill factor, $\alpha_{\textrm{\tiny meas}}$. Nominal fill factors were designed to be $\alpha_{\textrm{\tiny nom}}=0.016, 0.25\textrm{, and }0.77$ but due to over-etching fill factors were much larger. The lowest permittivity achieved was $\epsilon_{\textrm{\tiny eff}}=1.25$ with large triangular features. Generally, measured permittivity is lower than equation (1) predicts which we attribute to errors in de-embedding the sample-waveguide gap--a $50\mu$m change in the gap can shift permittivity by as much as 2.

We expected the fill factors to be nearly identical across geometries for a given feature size, but due to over-etching back-side features tend toward circular. At small feature sizes ($25\mu$m), where the features are over-etched due to longer etch time, this effect results in a widening of the square and triangular features at the end of the etch. This corresponds to significant undercut especially for the triangular features. The effect is far less pronounced for the small hexagons as they are already nearly circular--the resulting permittivity is higher. 

However, the effect appears to be reversed for the $100\mu$m and $175\mu$m feature sizes. Here, the total etch time is much smaller and results in minimal over-etch; larger features tend to etch faster and with greater resolution. The preference for circular etching is only barely present in medium features (slightly rounded edges) and completely absent in larger features. The dominant phenomenon is the tendency of larger features to etch faster; they experience greater undercut, larger fill factors and therefore lower permittivities.

We have fabricated triangle, square, and hexagon perforations of small ($25\mu$m), medium ($100\mu$m), and large ($175\mu$m) feature sizes on a $200\mu$m lattice in Silicon. We measured a minimum permittivity of $1.25$ which corresponds to $10\%$ of the background permittivity and enables much more flexibility when realizing TO designs and GRIN lenses. If we can perfect our small feature etching to account for over-etch we can more closely approach the maximum permittivity of $\epsilon_r=11.8$.







%

\bibliographystyle{IEEEtran}

\bibliography{aps_ursi_2017.bbl}

\begin{thebibliography}{1}
\providecommand{\url}[1]{#1}
\csname url@samestyle\endcsname
\providecommand{\newblock}{\relax}
\providecommand{\bibinfo}[2]{#2}
\providecommand{\BIBentrySTDinterwordspacing}{\spaceskip=0pt\relax}
\providecommand{\BIBentryALTinterwordstretchfactor}{4}
\providecommand{\BIBentryALTinterwordspacing}{\spaceskip=\fontdimen2\font plus
\BIBentryALTinterwordstretchfactor\fontdimen3\font minus
  \fontdimen4\font\relax}
\providecommand{\BIBforeignlanguage}[2]{{%
\expandafter\ifx\csname l@#1\endcsname\relax
\typeout{** WARNING: IEEEtran.bst: No hyphenation pattern has been}%
\typeout{** loaded for the language `#1'. Using the pattern for}%
\typeout{** the default language instead.}%
\else
\language=\csname l@#1\endcsname
\fi
#2}}
\providecommand{\BIBdecl}{\relax}
\BIBdecl

\bibitem{kwon_transformationEM_2010}
D.~H. Kwon and D.~H. Werner, ``Transformation electromagnetics: An overview of
  the theory and applications,'' \emph{IEEE Antennas and Propagation Magazine},
  vol.~52, no.~1, pp. 24--46, Feb 2010.

\bibitem{teichman_gradient_2013}
\BIBentryALTinterwordspacing
J.~Teichman, J.~Holzer, B.~Balko, B.~Fisher, and L.~Buckley, ``Gradient {Index}
  {Optics} at {DARPA},'' Institute for Defense Analyses, Tech. Rep. D-5027,
  Nov. 2013. [Online]. Available:
  \url{https://www.ida.org/~/media/Corporate/Files/Publications/IDA_Documents/STD/D-5027-FINAL.ashx}
\BIBentrySTDinterwordspacing

\bibitem{potosa_perforateddielectric_1994}
A.~Petosa and A.~Ittipiboon, ``Design and performance of a perforated
  dielectric fresnel lens,'' \emph{IEE Proceedings - Microwaves, Antennas and
  Propagation}, vol. 141, no.~5, pp. 309--, Oct 1994.

\bibitem{weir_material_1974}
W.~B. Weir, ``Automatic measurement of complex dielectric constant and
  permeability at microwave frequencies,'' \emph{Proceedings of the IEEE},
  vol.~62, no.~1, pp. 33--36, Jan 1974.

\end{thebibliography}




\end{document}